# Quantum Secure Biometric Authentication in Decentralised Systems

Tooba Qasim[1], Vasilios A. Siris[2], Izak Oosthuizen[3], Muttukrishnan Rajarajan[1], Sujit Biswas[1]
[1]City St. George's, University of London, UK
[2]Athens University of Economics and Business, Greece
[3]Zhero Cybersecurity and IT Support, London, UK
[1]{tooba.qasim, r.muttukrishnan, sujit.biswas}@citystgeorges.ac.uk,
[2]vsiris@aueb.gr, [3]io@zhero.co.uk

## Abstract

*Biometric authentication has become integral to digital identity systems, particularly in smart cities where it enables secure access to services across governance, transportation, and public infrastructure. Centralised architectures, though widely used, pose privacy and scalability challenges due to the aggregation of sensitive biometric data. Decentralised identity frameworks offer better data sovereignty and eliminate single points of failure but introduce new security concerns, particularly around mutual trust among distributed devices. In such environments, biometric sensors and verification agents must authenticate one another before sharing sensitive biometric data. Existing authentication schemes rely on classical public key infrastructure, which is increasingly susceptible to quantum attacks. This work addresses this gap by proposing a quantum-secure communication protocol for decentralised biometric systems, built upon an enhanced Quantum Key Distribution (QKD) system. The protocol incorporates quantum-resilient authentication at both the classical and quantum layers of QKD: post-quantum cryptography (PQC) is used to secure the classical channel, while authentication qubits verify the integrity of the quantum channel. Once trust is established, QKD generates symmetric keys for encrypting biometric data in transit. Qiskit-based simulations show a key generation rate of 15 bits/sec and 89% efficiency. This layered, quantum-resilient approach offers scalable, robust authentication for next-generation smart city infrastructures.*

## 1. Introduction

Biometric authentication has emerged as a fundamental pillar of modern identity verification systems, offering unique advantages over traditional password or token-based mechanisms [12]. From unlocking smartphones to verifying identities at national borders and enabling secure access to services like public transportation, digital governance, and smart healthcare systems in smart cities, biometric technologies such as facial recognition, fingerprints, iris scans, and voice patterns are now integral to both convenience and security. Their global adoption has surged in recent years. Governments, banks, and digital service providers increasingly depend on biometrics for trusted authentication, drawn by their resistance to loss, theft, and user error. Unlike passwords, biometric traits are inherently linked to individuals and thus more resilient against impersonation attempts [5].

However, the permanent nature of biometric data also brings serious risks. Unlike passwords, biometric information cannot be changed if it is stolen. For example, in 2015, a breach of U.S. government systems exposed 5.6 million fingerprint records, putting those individuals at risk of identity fraud for life [3]. Centralised systems that store biometric data on cloud servers or government databases are especially at risk. These centralised storage points are attractive targets for attackers, and if breached, they can lead to large-scale data leaks with lasting, irreversible consequences [26].

To reduce these risks, a shift toward decentralised architectures has begun. Approaches such as federated learning, verifiable credentials, and blockchain aim to minimise reliance on central servers by keeping biometric data closer to the user [4, 8]. While these strategies enhance privacy, they still depend on classical methods like RSA, Elliptic curve cryptography (ECC) and other public-key systems for secure communication, authentication, and key exchange among devices and entities in the biometric network. This dependence is a growing concern, as classical cryptographic methods are no longer considered future-proof in the face of emerging quantum computing capabilities.

The rise of quantum computing poses a serious threat to the cryptographic foundations underpinning these biometric systems. Algorithms such as RSA and ECC, which rely on the difficulty of factoring large integers or solving discrete logarithm problems, are expected to be broken by sufficiently powerful quantum computers using Shor's algorithm [22]. As a result, encrypted biometric templates and authentication mechanisms that appear secure today could be exposed in the near future, compromising user identities long after the data was initially shared or stored.

Biometric data, due to its permanence and sensitivity, is especially vulnerable in this context. If intercepted and decrypted, even years after its collection, it cannot be revoked or reset like a password. This makes biometric template protection not only a present-day concern but also a long-term security imperative. As noted in the ISO/IEC 24745 standard [10], effective biometric systems must ensure irreversibility, unlinkability, and renewability, properties that classical cryptographic protections may fail to uphold in the quantum era.

In response, the convergence of biometrics with quantum-safe solutions is gaining momentum. Post-quantum cryptography, built on hard mathematical problems such as lattice-based Learning With Errors (LWE), offers resistance against quantum adversaries and is already being standardised by NIST [17]. Meanwhile, QKD, grounded in the laws of quantum mechanics (e.g., no-cloning theorem, Heisenberg uncertainty principle), provides unconditional security for key exchange, enabling cryptographic protocols that are provably secure even against quantum-enabled adversaries [25]. In addition to QKD, Quantum Secure Direct Communication (QSDC) has emerged as a complementary paradigm that enables the direct transmission of secret messages over a quantum channel without requiring prior key exchange [18]. While not optimised for high-rate key generation, QSDC's principle of authentication between communicating parties serves as a foundational security layer in the proposed work.

This work is motivated by four major trends: (i) the shift from centralized to decentralized identity systems, creating new trust challenges. (ii) rising quantum threats to current authentication and encryption. (iii) increased use of biometrics in high-security environments and (iv) the difficulty of securely distributing symmetric keys across many devices in smart cities. To address these, we propose a layered, quantum-resilient trust model for securing biometric identity in decentralized systems, offering a scalable, future-ready solution aligned with smart city privacy and performance demands. The main contributions of this paper are:

1. **A quantum-secure communication framework** is proposed for decentralised biometric identity systems, combining Post-Quantum Cryptography and Quantum Key Distribution to safeguard biometric data transmission against quantum adversaries.

2. **A dual-layer authentication mechanism** is introduced within the QKD protocol, employing PQC on the classical channel and quantum authentication qubits on the quantum channel to ensure mutual trust between devices prior to key establishment.

3. **Extensive Qiskit-based simulations** validate the protocol's effectiveness, demonstrating high key generation rates and efficiency suitable for secure biometric communication in smart city deployments.

## 2. Background and Related Work

Biometric systems offer secure, convenient identity verification by using traits like fingerprints or facial features, which can't be lost or stolen like passwords. This makes them a powerful tool for protecting access to devices, services, and sensitive information. However, their growing use raises important privacy and data protection concerns.

The rise of quantum computing introduces new security challenges for biometric systems. Quantum algorithms like Shor's (1994) and Grover's (1996) have the potential to break widely used cryptographic methods that underpin many biometric authentication schemes. As a result, systems that rely on classical encryption techniques such as RSA, ECC or symmetric key cryptography may become vulnerable, threatening the integrity and security of biometric data in the quantum era [21, 23, 1].

The main concerns with biometric systems are protecting biometric templates and verifying the authenticity of transmitting systems. Templates are digital representations of permanent traits and if compromised, they can't be changed like passwords, posing a lifelong risk. In multi-user environments, ensuring the legitimacy of devices sending this data is equally critical. Researchers have proposed various template protection schemes, such as cancellable biometrics and biometric cryptosystems [20, 19, 24] but these schemes often rely on classical encryption techniques, such as RSA or ECC, that are vulnerable to emerging quantum attacks. Moreover, many of these schemes assume trusted environments or centralised architectures, which expose users to single points of failure and mass data breach risks [11].

While decentralised systems like federated learning and blockchain avoid single points of failure [14, 28, 6, 27], they still require the exchange of model updates and biometric templates, which are typically secured using classical cryptography. Federated learning allows collaborative model training without sharing raw biometric data, preserving user privacy. Blockchain ensures data integrity and traceability through its tamper-resistant ledger. While both enhance biometric security, they still rely on classical cryptography, which remains vulnerable to quantum attacks. To enhance privacy in biometric systems, homomorphic encryption (HE) has emerged as a promising technique, allowing computations on encrypted biometric templates without decryption, thus protecting sensitive data during authentica-tion [13]. However, many HE schemes, such as those based on Paillier [15] or Fan-Vercauteren methods [16], rely on classical cryptographic assumptions like integer factorisation or discrete logarithms, which are vulnerable to quantum attacks.

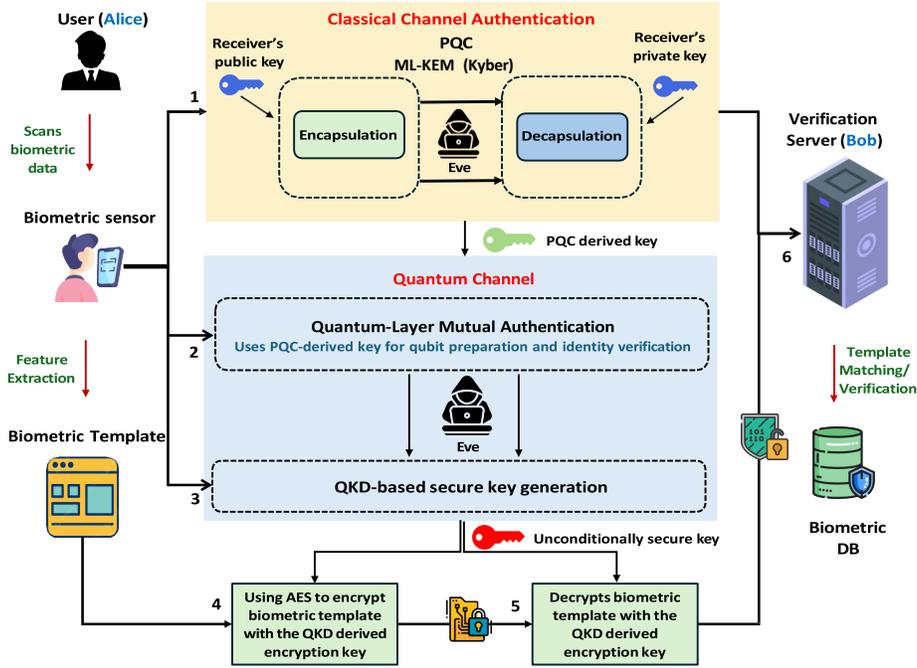

Figure 1. Secure biometric authentication workflow using PQC and QKD in smart city infrastructure.

Recent works propose lattice-based HE methods [9, 2] that offer some quantum resistance but remain computationally heavy, with complex key management and high overhead, making them impractical for real-time or IoT use [24]. Since they offer only computational security, future quantum advances could still break them. This highlights the need for a quantum-secure solution like QKD, which ensures provably unbreakable protection.

## 3. Proposed Protocol

### 3.1. Overview

The proposed protocol introduces a dual-channel authentication framework in QKD, aiming to address practical threats that arise from relying solely on classical authentication in biometric systems where devices such as sensors, edge agents, or terminals must securely exchange sensitive biometric templates or authentication signals. The workflow of the proposed protocol is shown in Fig. 1.

The protocol consists of two coordinated phases at both channels of QKD. In the first phase, PQC is used to establish classical authentication. Specifically, a lattice-based key encapsulation mechanism (ML-KEM) is employed to generate a shared secret between two entities, such as a biometric sensor (Alice) and a verification server (Bob). ML-KEM has been selected because of its proven security against quantum attacks and efficient key encapsulation with practical key sizes, making it well-suited for real-world, resource-constrained biometric systems. The generated secret is used to construct a sequence of authentication and decoy qubits as shown in Algorithm 1, which are transmitted over the quantum channel for mutual identity verification between the sender and receiver. Inspired by Quantum Secure Direct Communication, the authentication qubits confirm sender's identity, while decoy qubits are used to detect eavesdropping attempts through statistical channel monitoring. This ensures that only legitimate and trusted devices proceed with key establishment.

Both Alice and Bob authenticate each other using this sequence of authentication and decoy qubits and only after successful mutual authentication they proceed to the second phase, where a standard QKD process is used to transmit signal qubits and establish the final secret key. This staged approach ensures that key exchange is only conducted over a quantum channel that has already been authenticated, thereby strengthening protection against impersonation, injection, or relay attacks that may bypass classical verification, which preserves the privacy and authenticity of biometric data. These keys are subsequently used to encrypt biometric templates for secure transmission.

In the proposed framework, PQC is used solely for the initial authentication phase, eliminating the need for pre-shared keys, while QKD provides long-term, information-theoretic secure key generation. The PQC-derived key is not reused; instead, a portion of the first QKD session's key is used for symmetric-key authentication in subsequent rounds. This hybrid structure enables quantum-resilient mutual authentication and data confidentiality, reduces reliance on centralized trust anchors, and supports secure, scalable key management for practical biometric deployments in smart city environments.

## 3.2. Protocol Design and Operation

The proposed protocol involves four key steps: PQC-based classical authentication, where Alice and Bob establish a shared secret using ML-KEM; Quantum-layer authentication, where they verify each other's identity using authentication and decoy qubits; Signal qubit exchange via standard QKD to generate a secure key; and key reconciliation and privacy amplification to finalize a symmetric key for protecting biometric data.

### 3.2.1 Classical authentication using ML-KEM

Suppose Alice, a biometric sensor, wants to transmit a biometric template to Bob, a verification agent. Before initiating any exchange, both parties must authenticate each other. Alice and Bob begin by generating their respective ML-KEM public and private key pairs: ($Pub\_A$, $Priv\_A$) and ($Pub\_B$, $Priv\_B$). Alice randomly selects a secret and encapsulates it using Bob's public key $Pub\_B$, sending the ciphertext over the classical channel of QKD to Bob. Bob receives this ciphertext and decapsulates it using his private key $Priv\_B$ to recover the shared symmetric key. At this point, both parties hold an identical key, which is used to securely coordinate the quantum-layer authentication process that follows.

### 3.2.2 Authentication over quantum channel

After establishing a shared symmetric key, Alice uses it to prepare a sequence of authentication and decoy qubits based on Algorithm 1. This sequence is transmitted over the quantum channel to Bob. Upon receiving it, Bob measures the qubits using bases and positions based on Algorithm 2. The authentication qubits allow Bob to verify Alice's identity, while the decoy qubits serve as a statistical check for eavesdropping or tampering attempts on the quantum channel.

If the quantum bit error rate (QBER) observed from the authentication qubits remains below a defined threshold (3%), Bob considers the authentication successful. Otherwise, the session is aborted. For mutual authentication, Bob prepares and sends his own sequence of authentication and decoy qubits to Alice using the same procedure. Alice measures the received sequence and verifies Bob's identity. Only after both parties have successfully authenticated each other does the protocol advance to the key generation phase.

Table 1. Notations used in Algorithm 1 and 2

| Symbol | Description |
|---|---|
| $K$ | Pre-shared key used by Alice and Bob. |
| $K_d$ | Sequence of decoy qubits prepared by Alice. |
| $K_a$ | Sequence of authentication qubits prepared by Alice. |
| $K_A$ | Enlarged sequence formed by combining $K_d$ and $K_a$. |
| $k_{2i}$ | The $2i$-th bit of the pre-shared key $K$. |
| $k_{2i-1}$ | The $(2i-1)$-th bit of the pre-shared key $K$. |
| $k_{2i-1} \oplus k_{2i}$ | XOR operation between the $(2i-1)$-th and $2i$-th bits of $K$. |

### 3.2.3 Signal qubit exchange

Once mutual authentication is complete, Alice proceeds to transmit signal qubits to Bob using a standard QKD protocol. Bob measures the received signal qubits, and both parties perform basis sifting to extract a shared raw key. They then compute the QBER for the transmission. If the QBER exceeds a predefined threshold (pre-decided by Alice and Bob), the session is terminated to preserve security. Otherwise, they continue to the final phase of key reconciliation.

### 3.2.4 Key Reconciliation and Privacy Amplification

After successfully establishing a raw key, Alice and Bob perform error correction to resolve any discrepancies in their respective key strings. Once consistency is ensured, they apply privacy amplification techniques to remove any partial information that could have been gained by a potential eavesdropper, producing a final secure key. A portion of this key is reserved to authenticate future QKD rounds, while the rest is stored in a secure key pool for encrypting biometric templates and other sensitive communications. The sequence of decoy and authentication qubits generation is based on Algorithm 1, which is an extended version of the algorithm proposed in [7]. The original protocol introduced a theoretical qubit structure but lacked any practical simulation or performance evaluation. In this work, we modify that algorithm by incorporating a pre-shared key, established through ML-KEM based authentication to dynamically generate both decoy and authentication qubits. The modification also allows flexibility in controlling the ratio of decoy to authentication qubits. Increasing decoy qubits enhances eavesdropping detection, improving security in high-risk environments, while increasing authentication qubits reduces authentication time, optimizing the protocol for time-sensitive or low-risk scenarios. These enhancements not only enable practical deployment in real-world biometric systems but also allow the protocol to be tuned according to varying performance and security requirements. Simulation results validating the effectiveness of this approach are provided in Section 4.

Decoy qubits, randomly mixed with authentication qubits, statistically monitor the quantum channel for anomalies like photon number splitting (PNS) attacks. A high quantum bit error rate (QBER) from qubit measurements signals possible eavesdropping, prompting Alice and Bob to abort the session. To assess protocol performance, various decoy-to-authentication ratios are tested: 50/50, 30/70, and 10/90. These configurations help evaluate the trade-off and the protocol's reliability in verifying party identities.

**Algorithm 1** Preparation by Alice
　**Input:** Shared key $K$
　**Output:** Enlarged sequence $K_A$ sent to Bob
1: Generate Ordered Decoy Sequence $K_d$
2: **for** each $i$-th bit in $K$ **do**
3: 　**if** $k_{2i} = 0$ **then**
4: 　　Prepare the $i$-th qubit of $K_d$ in state $|0\rangle$ or $|1\rangle$
5: 　**else**
6: 　　Prepare the $i$-th qubit of $K_d$ in state $|+\rangle$ or $|-\rangle$
7: 　**end if**
8: **end for**
9: Generate Authentication Sequence $K_a$
10: **for** each $i$-th bit in $K$ **do**
11: 　**if** $k_{2i-1} \oplus k_{2i} = 0$ **then**
12: 　　Prepare the $i$-th particle of $K_a$ in state $|0\rangle$
13: 　**else**
14: 　　Prepare the $i$-th particle of $K_a$ in state $|-\rangle$
15: 　**end if**
16: **end for**
17: Form Enlarged Sequence $K_A$ by inserting authentication qubits into $K_d$
18: **for** each $i$-th bit in $K$ **do**
19: 　**if** $k_{2i-1} \oplus k_{2i} = 0$ **then**
20: 　　Place the $i$-th particle of $K_a$ after the $i$-th particle of $K_d$
21: 　**else**
22: 　　Place the $i$-th particle of $K_a$ before the $i$-th particle of $K_d$
23: 　**end if**
24: **end for**

**Algorithm 2** Measurement and Verification by Bob
　**Input:** Enlarged sequence $K_A$, shared key $K$
　**Output:** Authentication result and eavesdropping detection result
1: Determine position and basis for measuring decoy and authentication qubits using $K$
2: **for** each $i$-th qubit **do**
3: 　**if** $k_{2i} = 0$ **then**
4: 　　Measure using Z-basis $(|0\rangle, |1\rangle)$
5: 　**else**
6: 　　Measure using X-basis $(|+\rangle, |-\rangle)$
7: 　**end if**
8: **end for**
9: Use decoy qubits $K_d$ to analyze detection statistics and verify channel consistency
10: **if** anomalies detected in $K_d$ **then**
11: 　Abort communication
12: **else**
13: 　Use authentication qubits $K_a$ to calculate QBER
14: 　**if** QBER > 3% **then**
15: 　　Abort communication
16: 　**else**
17: 　　Authentication of Alice is successful
18: 　**end if**
19: **end if**

## 4. Evaluation and Analysis

### 4.1. Threat Model and Security Assumptions

This section outlines the key threats targeted by the proposed dual-channel authentication protocol in QKD.

**Defense Against MitM Attack:** In traditional QKD systems, a MitM attack may occur on the classical channel during public-key exchange or during post-processing stages. An adversary may impersonate a legitimate participant to intercept or alter communication. This protocol mitigates such attacks by employing the ML-KEM lattice-based PQC algorithm. Alice uses Bob's public key to encapsulate a secret:

$$(ct, ss_A) = \text{ML-KEM.Encapsulate}(pk_B)$$

where $ct$ is the ciphertext and $ss_A$ is the shared secret generated on Alice's side. Bob then decapsulates the received ciphertext $ct$ using his private key $sk_B$ to recover the shared secret $ss_B$:

$$ss_B = \text{ML-KEM.Decapsulate}(ct, sk_B)$$

If the shared secrets match, i.e., $ss_A = ss_B$, the authentication is successful, confirming the identities of both Alice and Bob, and ensuring the integrity of the communication.

This process effectively prevents MitM attacks. If an adversary (Eve) tries to intercept or modify the ciphertext $ct$, she would be unable to produce a valid shared secret without Bob's private key. Any tampering would result in mismatched shared secrets:

$$ss_A \ne ss_B$$

causing the session to be aborted. This prevents any unauthorized manipulation or impersonation during classical communication, even against quantum-capable adversaries.

**Defense Against PNS Attack:** PNS attacks exploit weak coherent pulses in practical QKD systems by splitting multi-photon states and retaining a copy without detection. To counter this, decoy qubits of varying photon intensities are inserted into the quantum stream. Bob analyzes detection statistics to identify inconsistencies. The probability $P_{\text{det}}$ of detecting an eavesdropper is given by:

$$P_{\text{det}} = 1 - \left(1 - \frac{n_d}{n_a + n_d}\right)^{n_e} \quad (1)$$

where $n_a$ is the number of authentication qubits, $n_d$ is the number of decoy qubits and $n_e$ is the number of qubits intercepted by the attacker. Equation 1 highlights the protocol's effectiveness in ensuring high detection probability. As the number of intercepted qubits increases or as the proportion of decoy qubits increases the probability of eavesdropper detection also increases.

***Impersonation at the Quantum layer:*** Even if classical authentication is secured, an attacker could inject forged quantum states to impersonate a party before classical verification completes. The proposed protocol includes a reciprocal quantum-layer authentication phase inspired by QSDC. Alice and Bob exchange authentication qubits encoded using a shared key. Each party verifies the expected measurement outcomes and computes the QBER. If QBER exceeds a threshold (3%), the session is aborted. Although the basis choices for authentication are deterministic which should result in 0% QBER in an ideal scenario but real-world quantum channels are subject to noise and physical imperfections. That is why a 3% threshold is set to allow for these minor imperfections while still detecting possible attacks. This quantum-layer identity verification protects against impersonation and injection attacks at the quantum level.

***Timing Based Side Channel Attacks:*** Hardware imperfections such as desynchronized photon sources can leak temporal patterns that reveal whether a qubit is a signal or decoy. These timing mismatches can be exploited by adversaries to infer qubit types and circumvent standard defenses. The proposed framework addresses this by authenticating the quantum channel before any signal qubits are transmitted. Since authentication occurs through a controlled sequence of qubits based on shared keys, any manipulation or timing inconsistency is flagged during the verification process.

## 4.2. Authentication Fidelity Analysis

To evaluate the fidelity of the quantum authentication phase in the proposed protocol, the QBER is analyzed. The focus is on how different decoy-to-authentication qubit ratios from Algorithm 1 affect anomaly detection and authentication reliability. In QKD, the QBER serves as a fundamental measurement of how effectively a protocol can transmit error-free bits over a quantum channel. It is a critical factor in determining the overall security and practicality of QKD systems.

For the proposed protocol, QBER was evaluated specifically during the quantum authentication phase. Mathematically QBER is defined as:

$$\text{QBER} = \frac{\text{Number of Incorrect Bits}}{\text{Total Number of Bits Transmitted}}$$

This ratio reflects the proportion of authentication bits that have been altered or corrupted during the transmission and measurement process. A QBER below 3% indicates high-fidelity authentication, essential for verifying communicating parties. Simulations with varying authentication-to-decoy qubit ratios (90%, 70%, and 50%) were conducted to assess their impact on performance as shown in Figure 2. Results show that larger key sizes help stabilise error rates, making detection more reliable. The ratio significantly affects performance: more authentication qubits speed up the authentication process but may slightly reduce the ability to catch anomalies, while more decoy qubits enhance detection but leaves fewer qubits for authentication. Through testing, a 70% authentication and 30% decoy split offered a good balance. This ratio can be adjusted depending on the security needs of the system.

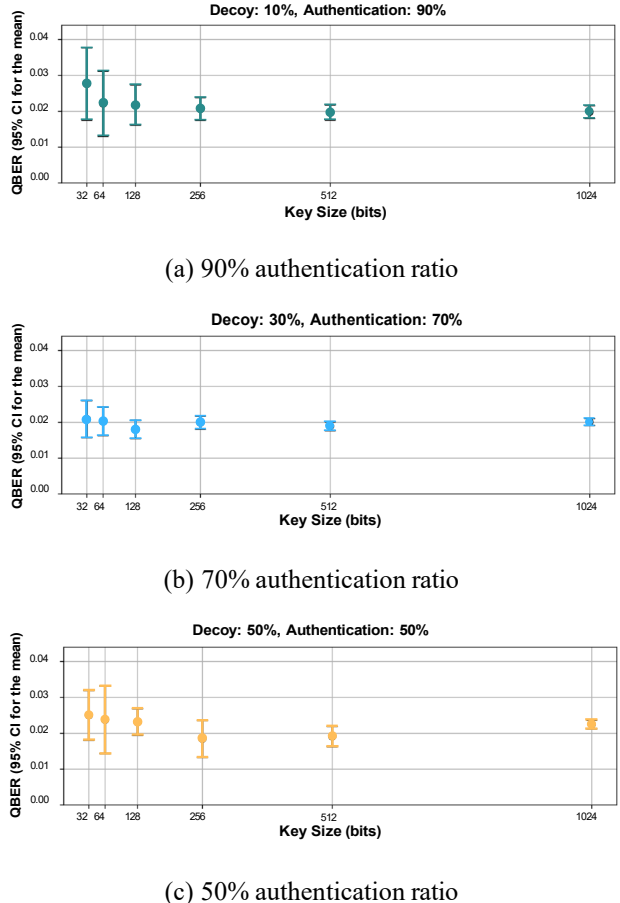

(a) 90% authentication ratio

(b) 70% authentication ratio

(c) 50% authentication ratio

Figure 2. QBER vs. key size under different authentication ratios.

## 4.3. Comparative Evaluation

To evaluate the effectiveness of the proposed dual-channel authentication protocol, it is compared with three well-established QKD protocols: SARG04, MDI-QKD and Twin Field QKD. These protocols are selected due to their practical significance and varied strengths in real-world

QKD deployments. All three rely on classical authentication schemes such as RSA or ECDSA, which are not secure against quantum adversaries. For consistency, ECDSA is used as the classical authentication method in the baseline implementations. SARG04 is included for its resilience to PNS attacks through modified sifting and error correction. MDI-QKD is chosen for its ability to eliminate detector-side vulnerabilities, and Twin Field QKD is selected for its high key rate performance over long distances, making it suitable for large-scale quantum networks.

In contrast, the proposed framework combines PQC with quantum-layer authentication, removing the need for pre-shared symmetric keys and enhancing resistance to both classical and quantum-layer attacks, including impersonation and injection. The comparative evaluation considers three key performance metrics: key generation rate, authentication overhead and protocol efficiency.

### 4.3.1 Testbed details

The simulations are performed on a 64-bit operating system with Windows 11 pro for Workstations installed, 32.0 GB RAM, and 1 TB storage capacity, providing sufficient resources for accurate testing and analysis. To design and execute the quantum circuits, we used Qiskit, IBM's open-source quantum computing framework, with each circuit run 100 times and measured using 1,000 shots.

### 4.3.2 Key Generation Rate

In QKD protocols, the key generation rate serves as a critical measure of efficiency. It quantifies the number of secure keys or bits generated per second, reflecting how quickly and effectively the protocol can produce usable keys for secure communication. As shown in Figure 3, the proposed protocol achieves a high key generation rate at low to moderate error rates, outperforming other established protocols under these conditions. However, at higher error rates, such as 40%, its performance decreases and converges with that of SARG04.

While all protocols experience a decline in key generation rate as error rates rise, reflecting the inherent limitations of QKD protocols in extremely noisy environments, the proposed protocol consistently achieves a higher rate compared to other protocols under low to moderate noise levels. SARG04, with its simpler structure, achieves a key generation rate comparable to the proposed protocol at higher error rates, but at the cost of reduced security. MDI-QKD also shows a decrease in key generation rate with rising errors, reflecting its added measures against detector-side attacks. TF-QKD, optimized for long-distance communication, has the lowest key generation rate, reflecting its focus on security over speed. The proposed protocol provides a good balance with enhanced security through PQC on the classical channel and quantum authentication, with a high key generation rate.

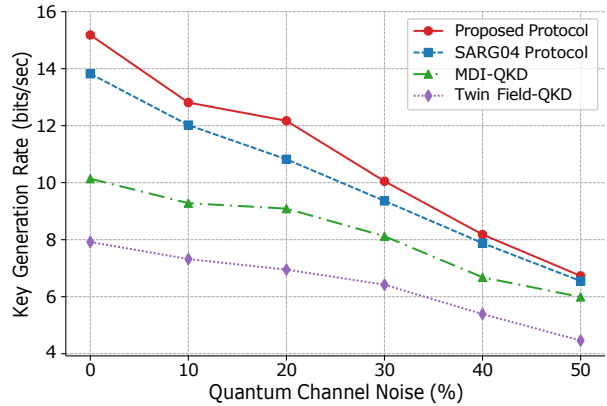

Figure 3. Key generation rate (bits per second).

Table 2. Runtime Analysis of the Proposed Dual-Channel Authentication Scheme

| Qubits Used | Auth. Time(s) | QKD Time(s) | Total Time(s) |
|---|---|---|---|
| 100 | 0.12 | 3.82 | 3.94 |
| 200 | 0.18 | 7.53 | 7.71 |
| 300 | 0.26 | 11.53 | 11.76 |
| 400 | 0.38 | 15.73 | 16.11 |
| 500 | 0.46 | 20.8 | 21.26 |
| 600 | 0.55 | 25.37 | 25.92 |
| 700 | 0.64 | 29.15 | 29.79 |
| 800 | 0.71 | 32.94 | 33.65 |
| 900 | 0.84 | 36.64 | 37.48 |
| 1000 | 0.96 | 41.52 | 42.48 |

### 4.3.3 Authentication Overhead

The overall runtime of the proposed dual-channel authentication scheme is compared with standard QKD protocols SARG04, MDI-QKD, TF-QKD, each using classical authentication mechanisms (ECDSA). Table 2 presents a detailed breakdown of the proposed protocol, showing the authentication time in seconds, which includes both classical and quantum authentication, QKD time and total session time for different numbers of qubits. Table 3 provides a comparison of the total time each protocol takes to complete, including both the authentication phase and the key distribution process.

Although the proposed approach adds some overhead from post-quantum cryptography and quantum-layer checks, it streamlines the QKD phase and ensures strong identity verification. A key advantage is that authentication occurs before QKD begins, avoiding wasted QKD attempts when authentication fails. Unlike other protocols that detect failure after QKD, this design conserves quantum resources and improves efficiency. Overall runtime remains competitive, especially compared to more complex methods like TF-QKD.

Table 3. Total Runtime Comparison Across QKD Protocols

| Qubits Used | Proposed (s) | SARG04 (s) | MDI-QKD (s) | TF-QKD (s) |
|---|---|---|---|---|
| 100 | 3.94 | 3.82 | 4.04 | 4.15 |
| 200 | 7.71 | 7.57 | 7.76 | 8.46 |
| 300 | 11.76 | 10.01 | 11.89 | 13.78 |
| 400 | 16.11 | 15.74 | 17.26 | 19.88 |
| 500 | 21.26 | 18.86 | 20.81 | 24.18 |
| 600 | 25.92 | 22.37 | 26.00 | 29.70 |
| 700 | 29.79 | 26.15 | 29.81 | 33.47 |
| 800 | 33.65 | 32.94 | 33.30 | 36.71 |
| 900 | 37.48 | 36.65 | 37.38 | 40.90 |
| 1000 | 42.48 | 40.52 | 41.69 | 44.67 |

### 4.3.4 Protocol Efficiency

Efficiency is a critical QKD metric, reflecting the percentage of raw key bits successfully converted into secure bits after processing. High efficiency (80–95%) indicates minimal bit loss during error correction and privacy amplification. As shown in Table 4, the proposed dual-channel authentication protocol achieves 89% efficiency, higher than MDI-QKD (48%) and Twin Field QKD (50%), and comparable to SARG04 (87%). This is due to its design, which performs quantum-layer authentication before signal qubit exchange. By initiating QKD only after mutual authentication, the protocol avoids unauthenticated or compromised sessions, reducing QBER and error rates during key generation. This results in more raw bits being retained as secure keys. Efficiency is influenced by QBER, channel noise, design, and distance, high QBER and noise lower it by increasing discarded bits. While MDI-QKD and Twin Field QKD offer strong security, their complex setups introduce higher overhead and reduce key retention. Thus, efficiency remains vital for balancing security and performance in QKD systems.

Table 4. Efficiency Comparison of QKD Protocols

| Protocol | Efficiency |
|---|---|
| SARG04 | 87% |
| MDI-QKD | 48% |
| Twin-Field | 50% |
| **Proposed Protocol** | **89%** |

### 4.4. Discussion

The simulation results show that the proposed dual-channel authentication protocol achieves strong security while maintaining practical performance. As demonstrated in Figure 3 and Table 4, it delivers a key generation rate of up to 15 bits/sec and an efficiency of 89%, outperforming MDI-QKD (48%) and Twin Field QKD (50%), and slightly exceeding SARG04 (87%). This high efficiency is due to the quantum-layer authentication step, which filters out unverified connections early and reduces wasted key material due to errors. Although the use of decoy and authentication qubits introduces additional steps, the runtime overhead remains minimal, making the proposed protocol efficient for practical use. As shown in Table 3, the runtime is only slightly higher than SARG04 and remains comparable to other traditional QKD methods. This slight overhead is a reasonable trade-off considering the added security gained through quantum-layer authentication. Algorithm 1 and 2 show how the insertion of decoy and authentication qubits at the quantum channel of QKD enables direct identity verification between Alice and Bob.

To assess adaptability, the protocol was tested with varying decoy-to-authentication qubit ratios, revealing a trade-off between security and response time. More decoy qubits improved eavesdropping detection, while more authentication qubits sped up identity verification time. This flexibility is valuable for biometric systems that require different security or response levels based on deployment needs. By combining classical PQC and quantum verification before key exchange, the protocol minimizes vulnerabilities and improves the trustworthiness of biometric data transfers. The layered structure ensures that only authenticated nodes participate in QKD, reducing error propagation and enhancing throughput in real-world, decentralized smart city deployments.

## 5. Conclusion and Future Work

This work presents a dual-channel authentication protocol for secure biometric communication in decentralized networks. It combines PQC-based classical authentication with quantum-layer verification to ensure only trusted devices engage in key exchange. Using ML-KEM and QSDC-inspired methods, it defends against impersonation, quantum threats, and channel tampering. Simulations show a key rate of 15 bits/sec and 89% efficiency with minimal runtime overhead. The protocol removes reliance on pre-shared keys and central authorities, making it suitable for scalable smart city deployments. Future work includes testing on real QKD hardware, integration with edge biometric systems, and scaling to multi-user mesh networks with dynamic trust requirements.

## 6. Acknowledgment

This work is supported by the TRUSTCHAIN EU project with the grant agreement number 101093274.